\documentclass[lengthcheck]{revtex4}
\usepackage{times,amsmath}
\usepackage{epsfig}
\begin{document}

\title{Optical Excitation of Nuclear Spin Coherence in Tm${}^{3+}$:YAG}
\author{A. Louchet, Y. Le~Du, F. Bretenaker,
T. Chaneli\`ere, F. Goldfarb, I. Lorger\'e and J.-L. Le~Gou\"et}

\affiliation{Laboratoire Aim\'e Cotton, CNRS-UPR 3321, Univ
Paris-Sud, B\^at. 505, 91405 Orsay cedex, France}
\author{O. Guillot-No\"el, Ph. Goldner}
\affiliation {Ecole Nationale Sup\'erieure de Chimie de Paris
(ENSCP), Laboratoire de Chimie de la Mati\`ere Condens\'ee de Paris,
CNRS-UMR 7574, ENSCP, 11 rue Pierre et Marie Curie 75231 Paris Cedex
05, France}


\begin{abstract}
A thulium-doped crystal is experimentally shown to be an excellent
candidate for broadband quantum storage in a solid-state medium. For
the first time, nuclear spin coherence is optically excited,
detected and characterized in such a crystal. The lifetime of the
spin coherence -- the potential storage entity -- is measured by
means of Raman echo to be about 300~$\mu$s over a wide range of
ground state splittings. This flexibility, attractive for broadband
operation, and well fitted to existing quantum sources, results from
the simple hyperfine structure, contrasting with Pr- and Eu- doped
crystals.
\end{abstract}
\maketitle

So far, the mapping of a quantum state of light onto an atomic
ensemble has been implemented in atomic vapors~\cite{julsgaard2001}
and cold atom clouds~\cite{kuzmich2003}. Because of their long
optical coherence lifetimes at low temperature, rare-earth ion-doped
crystals (REIC) have been extensively investigated for optical data
storage~\cite{mitsunaga1991} and data
processing~\cite{merkel1998,gorju2005}. There has recently been
renewed interest in these materials, stimulated by proposals to
explore their adequacy to quantum
memories~\cite{turukhin2002,longdell2005,alexander2006,hastings2006,guillotn2007}.
To some extent, REIC at low impurity concentration are similar to
atomic vapors with the advantage of no atomic diffusion.

Most optical quantum memory protocols rely on the transfer of a
quantum state of light into a long-lived atomic spin coherence that
is free from decoherence via spontaneous emission. This can be
achieved in a $\Lambda$-type three-level system where two hyperfine
or spin sublevels are optically connected to a common upper level.
The presence of a $\Lambda$-system ensures efficient coupling
between light and matter together with long storage times. A
$\Lambda$-system also represents a basic device where, in a simple
way, the transition to be excited by the quantum field can be
triggered for storage or restitution by an external control field.
In the prospect of quantum storage, electromagnetically induced
transparency (EIT) protocols involving $\Lambda$-type have been
studied in praseodymium-doped
materials~\cite{turukhin2002,longdell2005}.

It is noteworthy that the essence of $\Lambda$-system operation,
namely the optical excitation of nuclear spin coherence, has been
practised in REIC for almost 30
years~\cite{shelby1983,erickson1977,babbitt1989,blasberg1994,klieber2003a,ham1998},
but always in praseodymium- or europium-doped crystals. However,
quantum storage demonstration in Pr- or Eu-doped compounds is
limited by the smallness of their hyperfine structure, not really
matching the bandwidth of existing quantum sources. We recently
demonstrated the existence of a  $\Lambda$-system with adjustable
ground state splitting in thulium-doped YAG~\cite{deseze2006}.
Widely adjustable splitting might help to match the bandwidth of
existing quantum sources. In this Letter, for the first time to the
best of our knowledge, we optically excite, characterize and detect
the nuclear spin coherence in the electronic ground state of a
Tm-doped crystal.

In crystals doped with Pr${}^{3+}$ or Eu${}^{3+}$, $\Lambda$-systems
are built on the hyperfine structure of the ground level with a
sublevel splitting up to a few tens of MHz. This spacing represents
the memory bandwidth. Indeed, the two transitions of the $\Lambda$
cannot be polarization selected and thus only differ by their
frequency. Therefore, in order to excite only the relevant single
transition, the incoming signal must be spectrally narrower than the
splitting. Applying an external magnetic field would increase the
splitting and the memory bandwidth but would split each hyperfine
level into two nuclear spin sublevels, thus drastically
complexifying the level system. Besides, only dye lasers are
available at Pr${}^{3+}$ and Eu${}^{3+}$ operating wavelengths,
\textit{ie} 606~nm and 580~nm respectively in YSO. Because of the
high frequency noise generated by the dye jet, it is a challenging
task to achieve the sub-kHz linewidth sources that are needed  to
benefit from the long optical coherence lifetime. This limits the
realization to laboratory proof-of-principle. Although some groups
successfully built such laser
sources~\cite{sellars1994,klieber2003}, it is worth devising an
alternative approach.

The thulium rare earth ion actually gathers together many advantages
that make it specially attractive. First, its 793 nm wavelength
falls within reach of easily stabilized semiconductor lasers, unlike
Pr and Eu. Then, its $I=1/2$ nuclear spin gives rise, under external
magnetic field, to a simple straightforward 4-level optical system
in which a $\Lambda$-system can be selected. Moreover, because of
this simple structure, the sublevel splitting can be controlled
easily by the external magnetic field. This crystal could therefore
be used as an atomic quantum memory adapted to the bandwidth of
existing quantum sources. For these reasons, thulium-doped materials
are favorable alternatives to Pr- and Eu-doped compounds for quantum
storage applications, with the triple advantage of a tractable
wavelength, a simple level system and an adjustable ground state
splitting.

\begin{figure}
\includegraphics[height=3.5cm,angle=0]{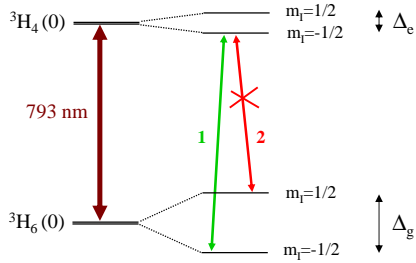}
\caption{(color online) Nuclear Zeeman effect in Tm:YAG. Nuclear
spin selection rule forbids transition 2.}\label{fig:lambda}
\end{figure}

Application of a magnetic field lifts the nuclear spin degeneracy,
but this is not enough to obtain a $\Lambda$-system (cf
Fig.~\ref{fig:lambda}). Indeed, optical excitation cannot flip a
nuclear spin, as expressed by a selection rule on the nuclear spin
projection $m_I$. However, coupling of electronic Zeeman effect and
hyperfine interaction enhances the nuclear Zeeman effect and gives
rise to nuclear gyromagnetic tensor anisotropy. If anisotropy is
different in ground and upper electronic levels, the nuclear spin
eigenvectors also differ in those two levels. As a consequence the
nuclear spin selection rule is relaxed. In YAG crystal, Tm ions are
doped into low symmetry sites ($D_2$) with a gyromagnetic tensor
anisotropy that is much larger in the electronic ground state than
in the excited state~\cite{deseze2006}. For an appropriate
orientation of the applied magnetic field, the optical transition
probability ratio along the two legs of the $\Lambda$ can be
optimized. In 2005, Guillot-No\"el \emph{et al.} theoretically
showed that, for an adequate external field orientation, the
branching ratio of the two transition probabilities reaches $0.24$
in 2 crystalline sites out of the 3 that can be selected by laser
beam polarization~\cite{guillotn2005}. In 2006, we directly measured
this optimal branching ratio to be $0.13\pm0.02$, proving that the
nuclear spin selection rule is effectively
relaxed~\cite{deseze2006,louchet2007branching}. An adjustable
$\Lambda$-system can now be built, involving both ground state
sublevels coupled to one of the two excited state sublevels.

\begin{figure}\centering
\includegraphics[width=8.6cm]{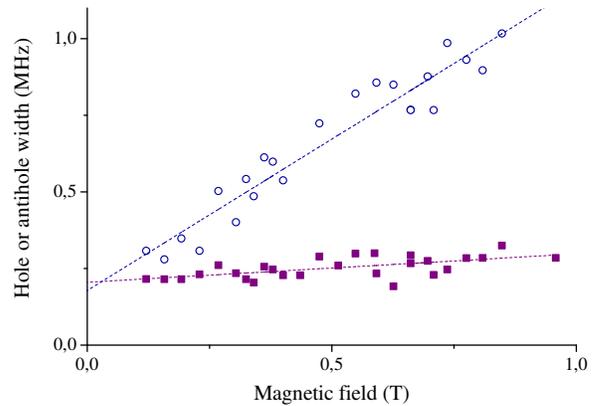}
\caption{(color online) Inhomogeneous broadening of side hole at
detuning $\Delta_e$ (filled squares) and of antiholes at detuning
$\Delta_g$ (empty circles). The $200$~kHz hole and antihole residual
width at zero field is assigned to chirped readout at
$3.5~10^{10}$~Hz/s. The dashed lines are fitted to the data. Their
slope is $0.99$~MHz/T for the antihole and $0.093$~MHz/T for the
hole.}\label{fig:inhw}
\end{figure}

The magnetic field $B$ being oriented to optimize the branching
ratio, the nuclear Zeeman sensitivity can be measured with the help
of hole burning spectroscopy. We find $\Delta_g/B = 36$~MHz/T and
$\Delta_e/B = 16$~MHz/T, where $\Delta_g$ and $\Delta_e$
respectively stand for the ground and excited electronic state
splittings. We also observe that, as depicted in
Fig.~\ref{fig:inhw}, the side hole and antihole widths exceed the
hole width at burning frequency, and vary linearly with the magnetic
field amplitude with a slope of $0.99$~MHz/T in the ground state,
and $0.093$~MHz/T in the excited state. We ascribe this broadening
to spatial variations of the magnetic field and substitution site
relative orientation. The external field orientation that optimizes
the transition probability branching ratio is close to the direction
of minimum splitting, and, accordingly, nearly orthogonal to the
main component of gyromagnetic tensor. This is the reason why we
measure a Zeeman sensitivity of $36$~MHz/T, although the main
gyromagnetic coefficient reaches $400$~MHz/T in the ground state. As
a consequence, a slight tilt of the field with respect to the site
frame entails dramatic splitting variation. At $1$~Tesla, a $1$~mrad
misorientation is enough to generate a $0.3$~MHz splitting deviation
in the ground state. At the moment it is not clear whether the
observed inhomogeneous broadening is caused by site orientation
disorder or by applied field orientational non-uniformity. The
ground state width appears to be 10 times more sensitive to the
magnetic field magnitude than the excited state width. Indeed the
gyromagnetic tensor is much less anisotropic in the upper level than
in the ground state. The ratio is consistent with an isotropic
misorientation model.

We turn now to investigating the spin coherence with optical means.
Initially the two ground state sublevels are equally populated and
the optical transition frequency is distributed over a $25$~GHz-wide
inhomogeneously broadened absorption profile. Within this huge
bandwidth, we select a narrow interval, smaller than
$\Delta_g-\Delta_e$, over which we prepare the ions for two-photon
excitation by pumping them into a single sublevel. Optical pumping
is accomplished by a sequence of $100~\mu$s chirped pulses. Then we
excite the ground state spin coherence with a 10~$\mu$s bichromatic
laser pulse. The two frequency components $\omega_1$ and $\omega_2$,
tuned to the $\Lambda$-system optical transitions, are separated by
$\Delta_g$ to achieve two-photon resonance. If all atoms remain
phased together, one could monitor the spin coherence evolution by
coherent forward Raman scattering: a long rectangular monochromatic
pulse at frequency $\omega_1$ excites one transition of the
$\Lambda$ and converts part of the spin coherence into an optical
coherence along the other transition. The resulting optical emission
could be detected as a beatnote at frequency $\Delta_g$ against the
probe pulse. Unfortunately, the spin coherences evolve at different
rates, according to the inhomogeneous broadening we observed in
Fig.~\ref{fig:inhw}. This makes the beatnote vanish on the timescale
of the excitation pulse duration. In  order to recover the optical
signature of the spin coherences, we resort to Raman echo
procedure~\cite{hartmann1968}. We apply an additional bichromatic
pulse, resonant with the two-photon transition, at mid-time between
initial excitation and final probing. This pulse reverses the spin
coherence time evolution, so that, at the moment of probing, the
spin coherences are phased together again and give rise to an
optical emission. The Raman echo signal is observed by means of
real-time Fast Fourier Transform.

\begin{figure}
\includegraphics[height=5.0cm,angle=0]{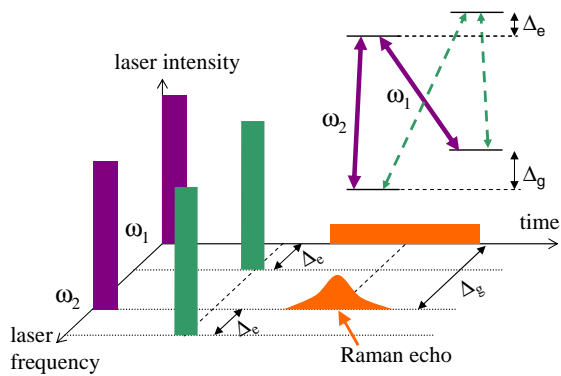}
\caption{(color online) Raman echo pulse sequence for photon echo
elimination using the two $\Lambda$-systems. The second pulse is
detuned from the first one by the excited state splitting $\Delta_e$
but still satisfies the two-photon resonance. Inset: 4-level system.
The first and second pulse frequencies $(\omega_1,\omega_2)$ and
$(\omega_1+\Delta_e,\omega_2+\Delta_e)$ are respectively depicted in
solid and dashed arrows.}\label{fig:double}
\end{figure}

The Raman echo is contaminated by two-pulse photon echo signals. The
photon echo signal at frequency $\omega_2$ beats with the detection
pulse, overlapping temporally and spectrally with the Raman echo. In
previous Raman echo experiments~\cite{ham1998}, photon echo was rejected by angular
separation of the various signals. In the present work we have
preferred a strictly collinear geometry that optimizes the spatial
mode matching of the various beams. Then, to get rid of photon echo
while maintaining an efficient driving of spin coherences, we detune
the two frequencies of the rephasing pulse so that they are
optically resonant with the other $\Lambda$-system of our 4-level
system (cf Fig.~\ref{fig:double}).

The system is illuminated with an extended cavity diode laser (ECDL)
operating at 793~nm, stabilized on a high-finesse Fabry-Perot cavity
through a Pound-Drever-Hall servoloop down to 200~Hz over
10~ms~\cite{crozatier2004}. The laser is amplified with a
semiconductor tapered amplifier (Toptica BoosTA). A polarizing cube
split the beam. Each component is double-passed through an
acousto-optic modulator (AOM) centered at 110MHz (AA
OptoElectronics). The two AOMs are driven by a dual-channel
1Gigasample/s waveform generator (Tektronix AWG520) that can provide
arbitrary amplitude and phase shaping. In this experiment each
channel feeds one frequency-shift at a time. After passing twice
through the AOMs, the split beams come back to the cube where they
merge into a fixed-direction single beam carrying the bichromatic
excitation. A common polarization direction is given by a Glan
prism. The recombined beam is finally coupled into a 2m-long
single-mode fiber. The light polarization direction is adjusted with
a half-wave plate to maximize the Rabi frequency. It is then focused
on the $5$~mm-thick, 0.1~at.\% Tm${}^{3+}$:YAG sample cooled down to
1.7~K in an Oxford Spectromag cryostat. The magnetic field generated
by superconducting coils is applied in the direction optimizing the
branching ratio~\cite{louchet2007branching}. The spot diameter on
the crystal is 80~$\mu$m. The transmitted light is collected on an
avalanche photodiode (Hamamatsu C5460 or C4777) protected from
strong excitation damaging light pulses by a third acousto-optic
modulator used as a shutter.

\begin{figure}
\includegraphics[height=5.5cm, angle=0]{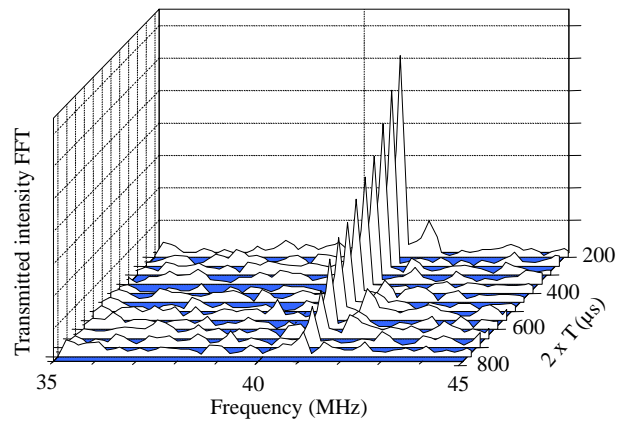}
\caption{(color online) Raman echo signals for $\Delta_g=41$~MHz as
delay time $T$ is increased from 100~$\mu$s to 400~$\mu$s. The spin
coherence lifetime is derived from the exponential decay of this
signal.}\label{fig:decay}
\end{figure}
\begin{figure}
\includegraphics[width=7cm]{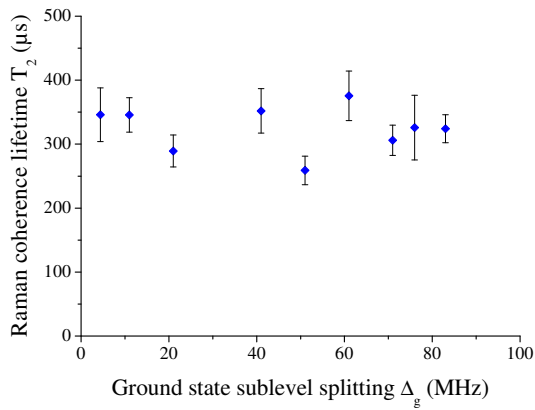}
\caption{Measurement of the spin coherence lifetime for different
ground state splittings, from $4$ to $83$~MHz. The error bars correspond to the standard error deduced from a least-squares fit.}\label{fig:T2}
\end{figure}

Raman echo experiments yield the spin coherence lifetime $T_2$. The
delay of the two bichromatic pulses is denoted $T$. The Raman signal
decays with $T$ as $e^{-2T/T_2}$. As an example,
Fig.~\ref{fig:decay} shows the Raman signal decay for a
$\Delta_g=41$~MHz ground state splitting. We measure the spin
coherence lifetime for different ground state splittings in
Tm${}^{3+}$:YAG. The low efficiency of the Raman echo process
accounts for the low signal to noise ratio. Thanks to the
double-pass setup, the two AOMs' $50$~MHz nominal bandwidth is
extended to $100$~MHz. However, $100$~MHz is the upper boundary not
for $\Delta_g$ but for $\Delta_g+\Delta_e\simeq 1.4 \Delta_g$, since
the excitation pulses contain not only frequencies $\omega_1$ and
$\omega_2=\omega_1-\Delta_g$, but also $\omega_1+\Delta_e$ and
$\omega_1-\Delta_g+\Delta_e$. At the limit of our equipment, we have
yet been able to reach a ground state splitting $\Delta_g$ up to
$83$~MHz. We show in Fig.~\ref{fig:T2} that the spin coherence
lifetime does not significantly depend on the ground state
splitting, remaining close to $300~\mu$s over a $80$~MHz range. The
error bars correspond to the standard error deduced from a
least-squares fit. We also measured the spin coherence lifetime in
the excited state to be $540\pm 35$$\mu$s for a $16.4$~MHz excited
state splitting with a similar Raman echo sequence. This lifetime is
close to the population lifetime of the upper electronic
state~\cite{macfarlane1993}. With a population lifetime of
$800~\mu$s, the intrinsic spin coherence lifetime turns out to
exceed $1$~ms. The mechanisms responsible for spin decoherence in
ground and excited states need to be unveiled by further studies.

Electromagnetically induced transparency (EIT) or controlled
reversible inhomogeneous broadening (CRIB)~\cite{moiseev2001}
experiments would be another step towards the demonstration of
thulium-doped crystals as quantum memories. In materials such as
ion-doped crystal, EIT might be regarded as a challenging operation.
Indeed, overcoming the inhomogeneous linewidth by the coupling field
Rabi frequency is known to be needed for efficient
EIT~\cite{kuznetsova2002}. However, atoms far from optical resonance
with the probe do not affect EIT, provided their ground state
sublevels are equally populated. Therefore the effective
inhomogeneous width to be considered essentially corresponds to the
population-unbalanced atoms close to optical resonance. Those atoms
cover a spectral interval much narrower than the inhomogeneously
broadened absorption line. In either Pr${}^{3+}$-, Eu${}^{3+}$- or
Tm${}^{3+}$-doped crystals this feature is essential to discard
non-resonant ions. In Eu${}^{3+}$- and Pr${}^{3+}$-doped crystals,
resonant atoms tend to be pumped to the third sublevel by the
driving fields and have to be repumped back to the selected
$\Lambda$ system ~\cite{turukhin2002,longdell2005} with the help of
an auxiliary beam. This preparation procedure only affects atoms
close to optical resonance. In Tm${}^{3+}$-doped crystals this
preparation step can probably be avoided since the ground state is
split in two sublevels only.

Linear Stark shift should be very small in Tm${}^{3+}$:YAG owing to
the $D_2$ site symmetry. Therefore inhomogeneous broadening is
difficult to reverse, which is not appropriate for CRIB.
Alternatively, one can consider to dope Tm ions into host matrices
exhibiting lower site symmetry such as Y${}_2$O${}_3$ or
LiNbO${}_3$. In these matrices, the Stark effect exists, allowing
the building of an artificial inhomogeneous broadening by
application of an electric field gradient. In LiNbO${}_3$, the
oscillator strength is found to be 50 times stronger than in
YAG~\cite{krishna2007}.

To conclude, for the first time we have optically excited, detected
and characterized a nuclear spin coherence in a thulium-doped
crystal. In addition to presenting a convenient absorption
wavelength and a simple and adjustable $\Lambda$-type 3-level
system, Tm${}^{3+}$:YAG also offers long ground state spin coherence
lifetimes, at splittings up to 80~MHz and probably much higher. This
lifetime stability over a wide frequency range proves the crystal
adequacy to existing quantum sources~\cite{shapiro2002,konig2005}.
The precise origin of the spin inhomogeneous broadening has still to
be clarified. Given its specific properties, thulium can be
considered as an excellent candidate for quantum storage in a solid,
challenging the previously studied praseodymium or europium.

\end{document}